\documentclass[a4paper,12pt]{article}
\pdfoutput=1 % if your are submitting a pdflatex (i.e. if you have
\usepackage{tikz}
\usetikzlibrary{arrows,snakes,backgrounds}
\usepackage{jheppub} % for details on the use of the package, please
\usepackage{amsfonts}
\usepackage{amssymb}
\usepackage{amsthm}
\usepackage{tikz}
\usepackage{subcaption}
\newcommand{\be}{\begin{equation}}
\newcommand{\eeq}{\end{equation}}
\newcommand{\bea}{\begin{eqnarray}}
\newcommand{\eea}{\end{eqnarray}}
\newcommand{\ba}{\begin{array}}
    \newcommand{\ea}{\end{array}}
\def\nn{\nonumber}

\newcommand{\ee}{\end{equation} }

\newcommand{\one}{{\rm 1\kern -.9mm l}}

\def\CD {{\cal D}}

\def\CF {{\cal F}}

\def\CM {{\cal M}}

\title{Correlation Functions of Classical and Quantum Artin System defined on Lobachevsky Plane and Scrambling Time}

\author{Hrachya Babujian,  Rubik Poghossian and  George Savvidy}

\affiliation{Institute of Nuclear and Particle Physics\\
 Demokritos National Research Centre,  Athens, Greece}
\emailAdd{babujian@yerphi.am,  poghos@yerphi.am,  savvidy@inp.demokritos.gr}

\affiliation{ Yerevan Physics Institute,\\
Alikhanian Br. 2, AM-0036 Yerevan, Armenia}

\abstract{We consider the quantisation of the Artin dynamical system defined on the fundamental region of the modular group.  In classical regime the geodesic flow in the fundamental region represents one of the most chaotic dynamical systems, it has mixing of all orders, Lebesgue spectrum and non-zero Kolmogorov entropy.  As a result, the classical correlation functions  decay exponentially. In order to investigate the influence of the classical chaotic behaviour on the quantum-mechanical properties of the Artin system we calculated the corresponding thermal quantum-mechanical correlation functions. It was conjectured by  Maldacena, Shenker and Stanford that the classical chaos can be diagnosed in thermal quantum systems by using an out-of-time-order correlation function as well as the square of the commutator of operators separated in time. We demonstrated that the two- and four-point correlation functions of the Louiville-like  operators decay exponentially with a temperature dependent exponent.  As conjectured the square of the commutator of the Louiville-like  operators  separated in time grows exponentially,  similar to the exponential divergency of trajectories in the classical regime. The corresponding exponent does not saturate the maximal growth condition.  }


\preprint{NRCPS-HE-26-2018, YerPhI/2018/45}

\begin{document}
    \maketitle
    \flushbottom

\section{\it Introduction}
 
The hyperbolic Anosov  C-systems  have exponential instability of their  trajectories and  as such  represent the most natural  chaotic dynamical systems \cite{anosov}. Of special interest are  C-systems which are defined on closed surfaces  of  the Lobachevsky plane of constant negative curvature.  An example of such system has been introduced in a brilliant article published in 1924 by the mathematician Emil Artin \cite{Artin}.  The  dynamical system is defined on the fundamental region of the Lobachevsky plane which is obtained by the identification of points  congruent with respect to the modular group $SL(2,Z)$,  a  discrete subgroup of the Lobachevsky plane isometries \cite{Poincare,Poincare1,Fuchs}.  The fundamental region $\CF$ in this case is a hyperbolic triangle on Fig.\ref{fig1}. The geodesic trajectories are bounded to propagate on the fundamental  hyperbolic  triangle.   The geodesic flow in this fundamental region represents one of the most chaotic dynamical systems with exponential instability  of its trajectories, has mixing of all orders, Lebesgue spectrum and non-zero Kolmogorov entropy 
\cite{hadamard,hedlund,anosov,bowen0,kolmo,kolmo1,sinai3,ruelle,hopf,Hopf,Gelfand,Collet,moore,dolgopyat,chernov,yangmillsmech,Savvidy:1982jk,yer1986a,konstantin,Savvidy:2015ida,Savvidy:2015jva,Maldacena:2015waa,Gur-Ari:2015rcq,Hanada:2017xrv}.

There is a great interest in considering quantisation of the hyperbolic dynamical systems and investigating their quantum-mechanical properties \cite{yangmillsmech,Savvidy:1982jk}. Our main interest in this article is a study of the behaviour of the correlation functions of the Artin hyperbolic dynamical system in its classical and quantum-mechanical regimes.   

In classical regime the correlation functions are defined as an integral over a pair of  functions/observables 
$A$ and $B $ in which the first one is stationary and the second one evolves with the 
geodesic flow $g_t$:
\bea
\CD_{t}(A,B) &=&   \int_{\CM} A(g) \overline{B(g g_t) } d \mu(g) .
\eea
The earlier investigation of the classical correlation functions of  the geodesic flows
was performed in 
\cite{Collet,Pollicot,moore,dolgopyat,chernov} by using different approaches including 
Fourier series for the $SL(2, R)$ group, zeta function for the geodesic flows, 
relating the poles of the Fourier transform of the correlation functions to the spectrum of an associated Ruelle operator,  the methods of unitary representation theory,  spectral properties of the 
corresponding Laplacian and other approaches.  In recent articles \cite{Savvidy:2018ygo,Poghosyan:2018efd} 
the authors  
demonstrated exponential decay of the correlation functions with time on the 
classical phase space.  The result was derived by using the differential geometry, group-theoretical  methods of Gelfand and Fomin, the time evolution equations and the properties of automorphic functions 
on $\CF$. The exponential decay rate was expressed in terms of the entropy $h(\CF)$ of the system: 
\bea
\vert \CD_t(A,B) \vert  & \leq &     ~ M~  e^{- K\,  h \,\vert t \vert}~~,
\eea
 where $M$ and $K$ are constants depending on the smoothness of the functions. 
{\it In classical regime the exponential divergency of the geodesic trajectories resulted into the universal exponential decay of its classical correlation functions} \cite{Savvidy:2018ygo,Poghosyan:2018efd}. 

In order to investigate the behaviour of the correlation functions in quantum-mechanical regime it is necessary  to know the spectrum of the system and the corresponding wave functions.  In the case of the modular group the energy spectrum has continuous part, which is originating from the asymptotically free motion inside an infinitely long "y -channel"  extended in the vertical direction of the fundamental region, as well as infinitely many discrete energy states  corresponding to a bounded motion at the "bottom"  of the fundamental triangle. The spectral problem has deep number-theoretical origin  and was partially solved in a series of pioneering articles \cite{maass,roeleke,selberg1,selberg2}.   It was solved partially because the discrete spectrum and the corresponding wave functions are not known analytically.   The general properties of the discrete spectrum have been derived by using Selberg trace formula \cite{selberg1,selberg2,bump, Faddeev,Faddeev1,hejhal2}.  Numerical calculation of the discrete energy levels were performed for many energy states   \cite{winkler,hejhal,hejhal1}.  

In the next section we shall describe the geometry of Lobachevsky hyperbolic plane and of the fundamental region which corresponds to the modular group $SL(2,Z)$, the geodesic flow on that region and the quantisation of the system.  The derivation of the Maass wave functions \cite{maass} for the continuous spectrum will be reviewed in details. We shall use  the Poincar\'e representation for Maass non-holomorphic automorphic wave functions. We introduce a natural physical variable $\tilde{y}$ for the distance in the vertical direction on the fundamental triangle  $\int dy/y =  \ln y = \tilde{y} $   and the corresponding momentum $p_y$ 
in order to represent the Maass wave functions (\ref{alterwave}) in the form which is appealing to the physical intuition 
\bea\label{alterwave0}
\psi_{p_y} (x,\tilde{y})  
= e^{-i p_y \tilde{y}  }+{\theta(\frac{1}{2} +i p_y) \over \theta(\frac{1}{2} -i p_y)}   \, e^{  i p_y \tilde{y}} + { 4  \over  \theta(\frac{1}{2} -i p_y)}   \sum_{l=1}^{\infty}\tau_{i p_y}(l)
K_{i p_y }(2 \pi  l e^{\tilde{y}} )\cos(2\pi l x).~~~~
\eea
Indeed, the first two terms describe the incoming and outgoing plane waves. The plane wave  $e^{-i p_y \tilde{y}  }$ incoming from infinity of the $y$ axis on Fig. \ref{fig1}  ( the vertex $\CD$)  elastically scatters on the boundary $ACB$ of the fundamental triangle  $\CF$ on Fig. \ref{fig1}.  The reflection amplitude is a pure phase and is given by the expression in front of the outgoing plane wave $e^{  i p_y \tilde{y}}$: 
\be\label{phase}
{\theta(\frac{1}{2} +i p_y) \over \theta(\frac{1}{2} -i p_y)} = \exp{[i\, \varphi(p_y)]}.
\ee
The rest of the wave function describes the standing waves $\cos(2\pi l x)$  in the $x$ direction between boundaries $x=\pm 1/2$
with the amplitudes $K_{i p_y }(2 \pi  l y )$, which are exponentially decreasing with index $l$ Fig.\ref{scattering}. The continuous energy spectrum is given by the formula  
\be
E=   p_y^2  + \frac{1}{4} .
\ee
The wave functions of the discrete spectrum have the following form \cite{maass,roeleke,selberg1,selberg2, winkler,hejhal,hejhal1}:
\bea\label{wavedisc0}
\psi_n(z) &=&   \sum_{l=1}^{\infty} c_l(n) \,
 K_{i u_n }(2 \pi  l e^{\tilde{y}} ) 
\{\begin{array}{ll} 
&\cos(2\pi l x) \\
&\sin(2\pi l x)   \\
\end{array}, 
\eea
where the spectrum $E_n = {1\over 4} + u^2_n$ and the coefficients $c_l(n)$  are not known analytically, but were computed numerically for many 
values of $n$ \cite{winkler,hejhal,hejhal1}.  

Having in hand the explicit expression of the wave function one can analyse a quantum-mechanical  behaviour of the correlation functions defined in \cite{Maldacena:2015waa}: 
\bea\label{basicopera1}
&\CD_2(\beta,t)=  \langle   A(t)   B(0) e^{-\beta H}   \rangle ,~~~\CD_4(\beta,t)=  \langle  A(t)   B(0) A(t)   B(0)e^{-\beta H}   \rangle \\
\label{basicopera2}
&C(\beta,t) =   -\langle [A(t),B(0)]^2 e^{-\beta H} \rangle~,
\eea
where in our case the operators $A$ and $B$ are chosen to be of  the Louiville type:
\be\label{basicopera}
A(N)=  e^{-2 N \tilde{y}},~~~N=1,2,.....
\ee
Analysing  the basic matrix elements of the Louiville-like  operators (\ref{basicopera}) we shall demonstrate   that all two- and four-point correlation functions (\ref{basicopera1}) decay exponentially with time, with the exponents which depend on temperature Fig.\ref{twopointfunc} and Fig.\ref{fourpointfunc}.  These exponents define the decorrelation time $t_d(\beta)$. 

Alternatively to the exponential decay of  the correlation functions (\ref{basicopera1}) the square of the commutator of the Louiville-like  operators  separated in time (\ref{basicopera2}) grows exponentially Fig.\ref{timeevolutionofcommutator}. This growth is reminiscent to the local exponential divergency of trajectories in the Artin system when it is considered in the classical regime \cite{Savvidy:2018ygo,Poghosyan:2018efd}. The exponential growth Fig.\ref{timeevolutionofcommutator} of the commutator (\ref{basicopera2}) does not saturate the condition of maximal growth (\ref{exponentc1}) and (\ref{linear1})
\be\label{exponentc12}
C(\beta,t) \sim f(\beta) \,e^{ {2\pi  \over  \chi(\beta)} t }, ~~~~\chi(\beta) \sim \beta
\ee
of the correlation functions which is conjectured to be linear in temperature   $T=1/\beta$
\be\label{exponentc1}
C(\beta,t) \sim f(\beta) \,e^{ {2\pi  \over  \beta} t }.
\ee
 
In our  calculation of the quantum-mechanical correlation functions we shall use  a perturbative expansion in which the high-mode Bessel's functions in (\ref{alterwave}) and (\ref{wavedisc}) will be  considered as perturbations.  We found that our calculations are stable with respect to these perturbations and do not influence the final results.  The reason is that in the integration region of the matrix elements (\ref{basicmatrix}) the high-mode Bessel's functions are exponentially small.

\section{\it Lobachevsky plane and its isometry group} 

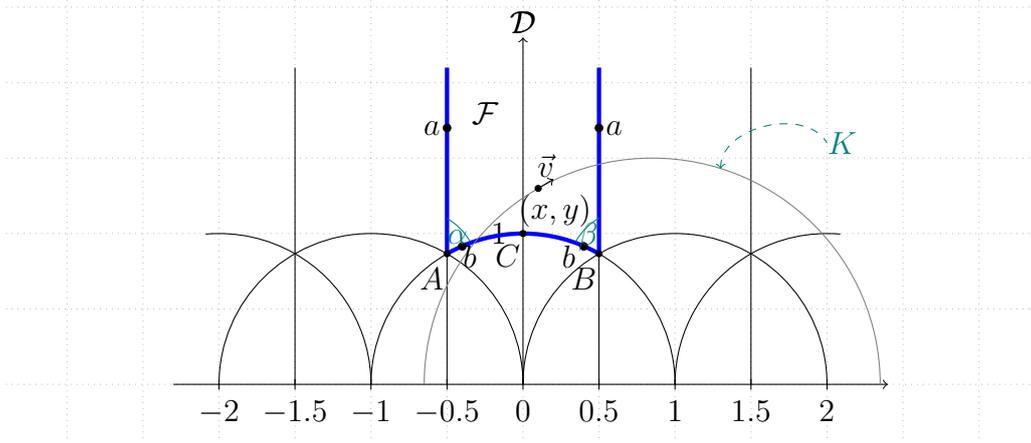
\begin{figure}
\hspace*{-1cm} 
\begin{tikzpicture}[scale=2]
\clip (-4.3,-0.4) rectangle (4.5,2.8);
\draw[step=.5cm,style=help lines,dotted] (-3.4,-1.4) grid (3.4,2.6);
\draw[,->] (-2.3,0) -- (2.4,0); \draw[->] (0,0) -- (0,2.3);
\foreach \x in {-2,-1.5,-1,-0.5,0,0.5,1,1.5,2}
\draw (\x cm,1pt) -- (\x cm,-1pt) node[anchor=north] {$\x$};
\foreach \y in {1}
\draw (1pt,\y cm) -- (-1pt,\y cm) node[anchor=east] {$\y$};
\draw (0,0) arc (0:180:1cm);
\draw (1,0) arc (0:180:1cm);
\draw (2,0) arc (0:180:1cm);
\draw (-1,0) arc (0:95:1cm);
\draw (1,0) arc (0:-95:-1cm);
\draw (1.5,0)--(1.5,2.1);
\draw (0.5,0)--(0.5,0.86602540378);
\draw (-0.5,0)--(-0.5,0.86602540378);
\draw (-1.5,0)--(-1.5,2.1);
\draw[ ultra thick, blue] (0.5,0.86602540378)--(0.5,2.1);
\draw[ ultra thick, blue] (-0.5,0.86602540378)--(-0.5,2.1);
\draw[ultra thick, blue] (0.5,0.86602540378) arc (60:120:1cm);
\draw (-0.25,1.8) node{ $\CF$};
\draw (0,2.4) node{ $\CD$};
\draw [thin,gray](2.35,0) arc (0:180:1.5cm);
\draw[fill] (0.1,1.29903810568) circle [radius=0.02];
\draw [->,] ((0.1,1.29903810568) to ((0.2,1.36);
\draw (0.21,1.15) node{ $(x,y)$};
\draw (0.15,1.45) node{ $\vec{v}$};
\draw (0,2.4) node{ $\CD$};
\draw (-0.6,0.7) node{ $A$};
\draw (0.4,0.7) node{ $B$};
\draw (-0.1,0.85) node{ $C$};
\draw[fill] (-0.5,0.86602540378) circle [radius=0.02];
\draw[fill] (0,1) circle [radius=0.02];
\draw[fill] (0.5,0.86602540378) circle [radius=0.02];
\draw[fill] (-0.5,1.7) circle [radius=0.025];
\draw[fill] (0.5,1.7) circle [radius=0.025];
\draw (-0.6,1.7) node{ $a$};
\draw (0.6,1.7) node{ $a$};
\draw[fill] ( -0.4,0.915) circle [radius=0.025];
\draw[fill] ( 0.4,0.915) circle [radius=0.025];
\draw (-0.35,0.85) node{ $b$};
\draw (0.3,0.85) node{ $b$};
\draw [<-,thin,  teal,dashed] (1.3,1.43) to [out=90,in=120] (2,1.6) ;
\draw  [thin,  teal] (2.1,1.6) node{ $K$};
\draw[thin,teal] (-0.5,1.1)  arc (60:30:0.41cm);
\draw[thin,teal] (0.35,0.952)  arc (150:120:0.41cm);
\draw [thin,teal] (-0.43,0.98) node{ $\alpha$};
\draw [thin,teal] (0.43,0.98) node{ $\beta$};
\end{tikzpicture}
\caption{The  non-compact fundamental region $\CF$ of a finite area is represented by the hyperbolic triangle $ABD$.  The vertex $D$ is at infinity of the $y$ axis and corresponds to a  cusp. The edges of the triangle are the arc $AB$,  the rays $AD$   and $BD$. The points on the edges $AD$ and $BD$ and the points of the arks $AC$ with $CB$ 
should be identified by the transformations $w=z+1$ and $w = -1/z$ in order to form a {\it closed  non-compact surface} $\bar{\CF}$
by  "gluing" the opposite edges of the modular triangle together.    The hyperbolic triangle $OAB$  can be considered equally well as the fundamental region. The modular transformations  of the fundamental  region $\CF$ create a regular tessellation of the  whole Lobachevsky plane by congruent hyperbolic triangles. 
 K is a geodesic trajectory passing through the point ($x,y$) of $\CF$ in the $\vec{v}$ direction.}
\label{fig1} 
\end{figure}

Let us start with Poincare model of
the Lobachevsky plane, i.e. the upper half of the complex plane:
$H$=$\{z \in \mathbb{C}$, $\Im z >0\}$ supplied with the metric (we
set $z=x+i y$) 
\bea 
dl^2=\frac{dx^2+dy^2}{ y^2}\, 
\label{metric_hp}
\eea 
with the Ricci scalar $R=-2 $. Isometries 
of this space are given by $SL(2,\mathbb{R})$ transformations. 
The $SL(2,\mathbb{R})$ matrix ($a$,$b$,$c$,$d$ are real and $ad-bc=1$ )
\[
g=\left(
\begin{array}{cc}
a&b\\c&d
\end{array}
\right)
\] 
acts on a point $z$ by linear fractional substitutions:
\bea
z\rightarrow \frac{az+b}{cz+d}~.
\eea
Note also that $g$ and $-g$ give the same transformation, hence 
the effective group is $SL(2,\mathbb{R})/\mathbb{Z}_2$.
We'll be interested in the space of orbits of a discrete subgroup
$G\subset SL(2,\mathbb{R})$ in $H$. Our main example will be 
the modular group $G=SL(2,\mathbb{Z})$. A nice choice of the 
fundamental region $\CF$ of $SL(2,\mathbb{Z})$ is displayed in Fig.\ref{fig1}.
The fundamental region $\CF$ of the  modular group consists of those  points between the lines
$x=-\frac{1}{2}$ and $x=+\frac{1}{2}$ that lie outside the unit circle in  Fig.\ref{fig1}.
The modular triangle $\CF$ has two 
equal angles  $\alpha = \beta = \frac{\pi }{3}$ and  with the third one equal to zero, $\gamma=0$, 
thus $\alpha + \beta + \gamma = 2 \pi /3 < \pi$.
The area  of the fundamental region is finite and equals to $\frac{\pi }{3}$ and gets a topology of sphere
by  "gluing" the opposite edges of the triangle. The invariant area element on the Lobachevsky plane is proportional to the square root of the determinant  of the matric (\ref{metric_hp}):
\be\label{me}
d \mu(z)= {dx dy \over y^2} \,.
\ee
thus
\be
\text{Area}(\CF)=\int _{-\frac{1}{2}}^{\frac{1}{2}} dx
\int _{\sqrt{1-x^2}}^{\infty }\frac{dy}{y^2}  =\frac{\pi }{3}\,.\nn
\ee

\section{\it Geodesic flow  in Hamiltonian gauge }

Consider geodesic flow on $\CF$, which is conveniently described 
by the least action principle $\delta S=0$, where (cf. with 
(\ref{metric_hp})):
\bea
S=\int L dt=\int \frac{\sqrt{\dot{x}^2+\dot{y}^2}}{ y}\,\,dt~.
\eea 
By varying the action, we immediately get the equations of motion
\bea
&&\frac{d}{dt}\,\, \frac{\dot{x}}{\,y\sqrt{\dot{x}^2+\dot{y}^2}}=0, 
\nonumber\\
&&\frac{d}{dt}\,\, \frac{\dot{y}}{\,y\sqrt{\dot{x}^2+\dot{y}^2}}
+\frac{\sqrt{\dot{x}^2+\dot{y}^2}}{\,y^2}=0.
\label{EOM_generic}
\eea 
 Notice the invariance of the action and of the equations under time reparametrizations 
$t\rightarrow t(\tau)$. Presence of a local ("gauge") symmetry 
indicates that we have a constrained dynamical system. 
One particularly convenient choice of gauge fixing
specifying the time parameter $t$ proportional  to the proper time, is archived by imposing 
the condition  
\bea
{ \dot{x}^2+\dot{y}^2 \over y^2}=2 H \,,
\label{gauge}
\eea
where $H$ is a constant. 
In this gauge the equations (\ref{EOM_generic}) will take the following form \cite{Faddeev:1969su}:
\bea
&&\frac{d}{dt}\,\,( \frac{\dot{x}}{\,y^2 })=0 
\nonumber\\
&&\frac{d}{dt}\,\, (\frac{\dot{y}}{\,y^2})
+\frac{2H }{\,y}=0.
\label{EOM_generic1}
\eea 
Defining the canonical momenta $p_x$, $p_y$ conjugate to the coordinates 
$x$, $y$  as 
\bea\label{basicequat}
p_x = \frac{\dot{x}}{\, y^2}\,, ~~~~
  p_y= 
\frac{\dot{y}}{\,y^2}~,
\eea
we shall get the geodesic equations   (\ref{EOM_generic1})  in the Hamiltonian form: 
 \bea
 \dot{ p_x}=0,~~~~
\dot{p_y} =-
\frac{2H }{\,y} .
\label{EOM_generic2}
\eea 
Indeed, after defining the Hamiltonian as
 \bea
 H = {1\over 2}y^2 (p^{2}_{x} +p^{2}_{y})
\label{hamiltonian1}
\eea 
the corresponding equations will take the following form: 
\bea
&& \dot{x} = \frac{\partial H }{\, \partial p_x} = y^2 p_x,~~~~~~\dot{y} = \frac{\partial H }{\, \partial p_y} = y^2 p_y
\nonumber\\
&&\dot{p_x} = - \frac{\partial H }{\, \partial x} = 0,~~~~~~~~\dot{p_y} = -\frac{\partial H }{\, \partial y} = - y(p^2_x +p^2_y)= -\frac{2H }{\,y}
\eea
and coincide with (\ref{basicequat}) and (\ref{EOM_generic2}).
The advantage of the gauge (\ref{gauge}) is that the Hamiltonian (\ref{hamiltonian1}) coincides with the constraint. 
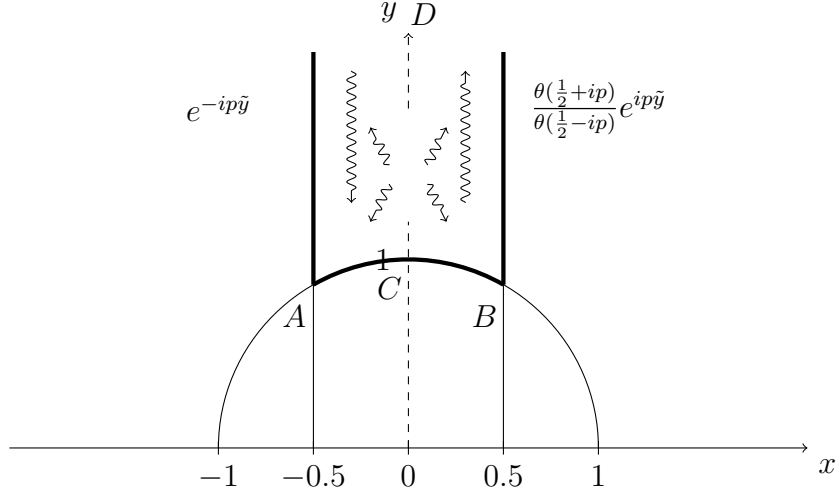
\begin{figure}[htbp]
	 \hspace*{-3.5cm} 
	\begin{tikzpicture}[scale=2.5]
	\clip (-4.3,-0.4) rectangle (4.5,2.8);
	\draw[,->] (-2.1,0) -- (2.1,0)  node[anchor=north west] {$x$ };
	\draw[dashed] (0,0) -- (0,1.2);
	\draw[dashed,->]  (0,1.8)--(0,2.2)  node[anchor=south east] {$y$};
	\foreach \x in {-1,-0.5 ,0,0.5,1}
	\draw (\x cm,1pt) -- (\x cm,-1pt) node[anchor=north] {$\x$};
	\foreach \y in {1}
	\draw (1pt,\y cm) -- (-1pt,\y cm) node[anchor=east] {$\y$};
	\draw (1,0) arc (0:180:1cm);
	\draw (0.5,0)--(0.5,0.86602540378);
	\draw (-0.5,0)--(-0.5,0.86602540378);
	\draw[ ultra thick,black] (0.5,0.86602540378)--(0.5,2.1);
	\draw[ ultra thick, black] (-0.5,0.86602540378)--(-0.5,2.1);
	\draw[ultra thick, black] (0.5,0.86602540378) arc (60:120:1cm);
	\draw (-0.6,0.7) node{ $A$};
	\draw (0.4,0.7) node{ $B$};
	\draw (-0.1,0.85) node{ $C$};
	\draw [->,snake=snake,
	segment amplitude=.6mm,
	segment length=1.5mm,
	line after snake=1mm] (-0.3,2) -- (-0.3,1.3);
	\draw [->,snake=snake,
	segment amplitude=.6mm,
	segment length=1.5mm,
	line after snake=1mm] (0.3,1.3)-- (0.3,2);
	%%%%%%%%%%%%%%%%%%%%%%%%%%%%%%%%%%%%%%%%%%%%%%%%%%%%
	\draw [->,snake=snake,
	segment amplitude=.6mm,
	segment length=1.5mm,
	line after snake=1mm] (0.1,1.5) -- (0.2,1.7);
	%%%%%%%%%%%%%%%%%%%%%%%%%%%%%%%%%%%%%%%%%%%%%%%%%%%
	\draw [->,snake=snake,
	segment amplitude=.6mm,
	segment length=1.5mm,
	line after snake=1mm] (-0.1,1.5) -- (-0.2,1.7);
	%%%%%%%%%%%%%%%%%%%%%%%%%%%%%%%%%%%%%%%%%%%%%%%%%%%
	%%%%%%%%%%%%%%%%%%%%%%%%%%%%%%%%%%%%%%%%%%%%%%%%%%%
	\draw [->,snake=snake,
	segment amplitude=.6mm,
	segment length=1.5mm,
	line after snake=1mm] (-0.1,1.4) -- (-0.2,1.2);
	%%%%%%%%%%%%%%%%%%%%%%%%%%%%%%%%%%%%%%%%%%%%%%%%%%%
	%%%%%%%%%%%%%%%%%%%%%%%%%%%%%%%%%%%%%%%%%%%%%%%%%%%
	\draw [->,snake=snake,
	segment amplitude=.6mm,
	segment length=1.5mm,
	line after snake=1mm] (0.1,1.4) -- (0.2,1.2);
	%%%%%%%%%%%%%%%%%%%%%%%%%%%%%%%%%%%%%%%%%%%%%%%%%%%
	\draw (1,1.8) node{ $\frac{\theta({1\over 2}+i p)}{\theta({1\over 2}-i p)}e^{i p \tilde{y}}$};
	\draw (-1,1.8) node{ $e^{-i p \tilde{y}}$};
	%%%%%%%%%%%%%%%%%%%%%%%%%%%%%%%%%%%%%%%%%%%%%%%%%%%%%%%
	\draw (0.08,2.3) node{ $D$};
	\end{tikzpicture}
 
	\caption{ The incoming and outgoing plane waves. The plane wave  $e^{-i p \tilde{y}  }$ incoming from infinity of the $y$ axis on Fig. \ref{fig1}  ( the vertex $\CD$)  elastically scatters on the boundary $ACB$ of the fundamental triangle  $\CF$ on Fig. \ref{fig1}.  The reflection amplitude is a pure phase and is given by the expression in front of the outgoing plane wave $e^{  i p \tilde{y}}$.
The rest of the wave function describes the standing waves in the $x$ direction between boundaries $x=\pm 1/2$ with the amplitudes, which are exponentially decreasing.}
	
	\label{scattering}
\end{figure}

\section{\it Quantization}

Now it is fairly standard to quantize this Hamiltonian system. 
We simply replace in (\ref{hamiltonian1})   
\[
p_x=-i\frac{\partial}{\partial x}, \qquad 
p_y=-i\frac{\partial}{\partial y}
\]
and consider the (time independent) Schr\"odinger equation
\[
H\psi = E \psi.
\]
The resulting equation explicitly reads:
\bea
-y^2(\partial_x^2+\partial_y^2)\psi= E \psi.
\eea 
On the lhs one easily recognises  the  Laplace operator (with an 
extra minus sign) in Poincare metric (\ref{metric_hp}).
It is easy to see that the Hamiltonian is positive semi-definite Hermitian operator.
Indeed, for any quadratically integrable function $\psi(x,y)$
 \bea
-\int \psi^*(x,y )\, y^2(\partial_x^2+\partial_y^2) \, \psi(x,y ) {d x dy \over y^2}=  \int (\vert \partial_x \psi(x,y)\vert^2+ \vert \partial_y \psi(x,y)\vert^2) d x dy  \geq 0.~~~~
\label{ semiposti}.
\eea 
It is convenient to introduce parametrization of the energy $E=s(1-s)$ and 
to rewrite this equation as
\bea
-y^2(\partial_x^2+\partial_y^2)~ \psi(x,y)  = s(1-s) ~\psi(x,y).
\label{Laplace_eq}
\eea 
As far as $E$ is real and semi-positive and parametrisation is symmetric with respect to $s \leftrightarrow 1-s $ it follows that the parameter $s$ should be chosen within the range
\be\label{srange}
s \in  [1/2 , 1  ]~~ \text{or } ~~ s=1/2 +i u ,~~~u ~\in~ [0,\infty].
\ee
One should impose the "periodic" boundary condition on the wave function  with respect to the modular group 
\bea\label{invariance}
\psi(\frac{a z+b}{c z+d})=\psi(z),~~~\left(
\begin{array}{cc}
a&b\\c&d
\end{array}
\right) \in SL(2,Z)
\eea
in order to have  the wave function which is properly defined on the fundamental 
region $\bar{\CF}$ shown in Fig. \ref{fig1} .
Taking into account that the transformation $T:z\rightarrow z+1$ 
belongs to $SL(2,Z)$, one has to impose the periodicity condition 
$\psi(z)=\psi(z+1)$. Thus we have a Fourier expansion
\bea
\psi(x,y)=\sum_{n=-\infty}^\infty f_n(y)\exp(2\pi i n x).
\eea  
Inserting this into Eq. (\ref{Laplace_eq}), for the Fourier 
component $f_n(y)$ we get
\bea
\frac{d^2f_n(y)}{dy^2}+(s(1-s)-4\pi^2n^2)f_n(y)=0~.
\eea   
For the case $n\neq 0$ the solution which exponentially decays at 
large $y$ reads  
\bea
f_n(y)=\sqrt{y} K_{s-\frac{1}{2}}(2\pi n|y|),
\eea
and for $n=0$ one simply gets
\bea\label{partsolution}
f_0(y)=c_0 y^s+c^{'}_0 y^{1-s}.
\eea 
Thus the solution can be represented in the following form: 
\bea\label{solution3}
\psi(x,y) =  c_0 y^s+c^{'}_0 y^{1-s} + \sqrt{y}\sum_{n=-\infty \atop n \neq 0}^\infty c_n  K_{s-\frac{1}{2}}(2\pi n|y|) \exp(2\pi i n x),
\eea
where the coefficients $c_0, c^{'}_0, c_n$ should be defined so that the wave function fulfils  the boundary conditions (\ref{invariance}). Thus one should  impose also the invariance with respect to the second generator of the modular group $SL(2,Z)$, that is,  with respect to the  transformation $S:z\rightarrow -1/z$ ~:
\bea
\psi(z)=\psi(-1/z).
\eea
This functional equation defines the coefficients $c_0, c^{'}_0, c_n$.  We found that it is much easier to resolve it by using the full group of $SL(2,Z)$ transformations acting on a particular solution (\ref{partsolution}).  The wave function generated  in this way will be  invariant with respect to the $SL(2,Z)$ transformations. We shall follow this approach in the next section. 

\section{\it Continuous spectrum and the reflection amplitude}

As we just mentioned above in order to get $SL(2,Z)$ invariant solutions, one should define the coefficients $c_0, c^{'}_0$ and $c_n$ in (\ref{solution3}).   
Another option is to start from a particular solution and 
perform summation over all nonequivalent shifts of the argument
by the elements of $SL(2,Z)$, that is, by using the Poincar\'e  series representation \cite{Poincare,Poincare1}. 
We'll demonstrate this strategy 
by using  the simplest solution (\ref{partsolution}), (\ref{solution3}) with $c_0 =1, c^{'}_0=0$:
\[
\psi(z)= y^s = (\Im z)^s\, .
\] 
Let us denote by $\Gamma_{\infty}$ the
subgroup of $\Gamma = SL(2,Z)$, generating shifts $z\rightarrow z+n$, 
$n\in Z$. Explicitly the elements of $\Gamma_{\infty}$ are given by 
$2\times 2$ matrices:
\bea
g_n=\left(
\begin{array}{ccc}
1&n\\
0&1
\end{array}
\right).
\eea
Since $y^s$ is already invariant with respect to $\Gamma_{\infty}$, 
we should perform summation over the conjugacy classes 
$\Gamma_{\infty} \backslash \Gamma $. Let us define these conjugacy classes. If two $SL(2,Z)$ matrices 
\[
\left(
\begin{array}{ccc}
a&b\\
c&d
\end{array}
\right) 
\qquad {\text and} \qquad
\left(
\begin{array}{ccc}
a'&b'\\
c'&d'
\end{array}
\right)
\]
belong to the same class, then by definition for some $n\in Z$
\[
\left(
\begin{array}{ccc}
a'&b'\\
c'&d'
\end{array}
\right) 
=
\left(
\begin{array}{ccc}
1&n\\
0&1
\end{array}
\right)
\left(
\begin{array}{ccc}
a&b\\
c&d
\end{array}
\right)
\]
so that $c'=c$, $d'=d$, $a'-a=nc$ and $b'-b=nd$. Since $ad-bc=1$, it follows  that 
$a$ and $c$ do not have a common devisor. In fact, the opposite is 
also true. Given a pair of mutually prime integers $(c,d)$ it is 
always possible to find a pair of integers $(a,b)$ such that 
$ad-bc=1$.  For any other pair $(a',b')$  satisfying the same condition $a'd-b'c=1$, 
the relations $a'-a=nc$ and $b'-b=nd$ are satisfied  for some integer $n$. 
Thus we established a bijection between the set of mutually 
prime pairs  $(c,d)$ with $(c,d)\neq (0,0)$ and the set of conjugacy 
classes $\Gamma_\infty\backslash \Gamma $. The fact that the integers
$(c,d)$ are mutually prime integers means that their greatest common divisor 
(gcd) is equal to one: $gcd(c,d)=1$.  As a result, it is defined by the classical Poincar\'e  series representation \cite{Poincare,Poincare1} and  
for the sum of our interest we get\footnote{ The factor $1/2$ below is introduced 
for removing the double degeneracy due to the fact that $SL(2,Z)$ 
elements $\pm \gamma $ both act on $z$ in the same way.}
\bea
\psi_{s}(z) \equiv \frac{1}{2}\sum_{\gamma \in \Gamma_\infty\backslash \Gamma}
(\Im (\gamma z))^s=\frac{1}{2}\sum_{(c,d) \in \mathbb{Z}^2 \atop gcd(c,d)=1} \frac{y^s}{((c x+d)^2+c^2y^2)^s}~,
\label{sum1}
\eea   
where, as explained above, the sum on r.h.s. is taken over all 
mutually prime pairs $(c,d)$. The series (\ref{sum1}) is convergent when $\Re\, s>1$. We used also the simple 
relation
\[
\Im \gamma z\equiv \Im \frac{az+b}{cz+d}=
\frac{y}{(c x+d)^2+c^2y^2}~.
\] 
To simplify further the sum let us multiply  both sides of the
eq. (\ref{sum1}) by \cite{maass}
\[
\sum_{n=1}^{\infty}\frac{1}{n^{2s}}
\equiv \zeta(2s)
\]  
so that we shall get
\bea
\zeta(2s) \, \psi_{s}(z) = \frac{1}{2}\sum_{n=1}^\infty
\sum_{(c,d) \in \mathbb{Z}^2 \atop gcd(c,d)=1} 
\frac{y^s}{((nc x+nd)^2+(nc)^2y^2)^s} .
\eea 
It is easy to get convinced now that the set of all pairs 
$(nc,nd)$ with $n$ a positive integer and $(c,d)$ - mutually prime, 
coincides with the set of all pairs of integers $(m,k)$ 
which are not simultaneously zero. Indeed, given a pair $(m,k)$    
we can factor out the greatest common divisor $n$ and represent it 
as $(nc,nd)$ with  mutually prime $(c,d)$.
Thus we arrive at the Eisenstein series representation of the wave function: 
\bea\label{solution4}
\zeta(2s) \, \psi_{s}(z) =\frac{1}{2} 
\sum_{(m,k) \in \mathbb{Z}^2 \atop (m,k)\neq(0,0)}
\frac{y^s}{((m x+k)^2+m^2y^2)^s} .
\eea
Since the r.h.s. of this equation is periodic in $x$ with period $1$, 
we can expand it in Fourier series. Our next goal is to find the 
coefficients of this expansion:
\bea\label{fcoefficients}
c_l(y)=\frac{1}{2} 
\sum_{(m,k)\in \mathbb{Z}^2\atop (m,k)\neq(0,0)}
\int_0^1 \frac{y^se^{-2\pi il x}dx}{((m x+k)^2+m^2y^2)^s} .
\eea
First let us handle the term with $m=0$:
\bea 
\frac{1}{2} 
\sum_{k\in \mathbb{Z}\atop k\neq0 }
\int_0^1 \frac{y^se^{-2\pi i l x}dx}{k^{2s}}=\delta_{l,0}\zeta(2s)y^s .
\eea 
For the sum over non-zero $m$'s let's notice that we may drop the factor 
$1/2$ and sum over $m\ge 1$. Indeed, the sum over negative $m$'s can 
be reverted to a sum over positive ones through redefinition
$k\rightarrow -k$. For fixed positive $m$ it is instructive to 
represent $k$ as $k=n m+s$ thus splitting the initial sum over $k\in \mathbb{Z}$ into double sum over $r=0,1,\ldots,m-1$ and $n\in 
\mathbb{Z}$. In this way after few simple manipulations we get
\bea  
&&\sum_{m=1}^\infty\sum_{r=0}^{m-1}\sum_{n\in \mathbb{Z}}
\int_0^1 \frac{y^se^{-2\pi il x}dx}{((m (x+n)+r)^2+m^2y^2)^s}
=\sum_{m=1}^\infty\sum_{r=0}^{m-1}
\int_{-\infty}^{\infty} \frac{y^se^{-2\pi il x}dx}
{((m x+r)^2+m^2y^2)^s}\nonumber\\
&&=\sum_{m=1}^\infty\sum_{r=0}^{m-1}m^{-2s}y^{1-s}
e^{\frac{2\pi ilr}{m} }\int_{-\infty}^{\infty} 
\frac{\cos(2\pi |l|yx)dx}{(x^2+1)^s} .
\eea
The last integral is expressed in terms of modified Bessel's 
$K$ function:
\bea
\int_{-\infty}^{\infty} 
\frac{\cos(2\pi |l|yx)dx}{(x^2+1)^s}=\left\{
\begin{array}{l}
\frac{2 \pi ^s} {\Gamma (s)} 
\left|ly\right|^{s-\frac{1}{2}}
K_{s-\frac{1}{2}}(2 \pi  |l|y ), 
\quad {\text if}\,\, l\neq 0\\
\frac{\sqrt{\pi } \Gamma \left(s-\frac{1}{2}\right)}{\Gamma (s)},
\quad \quad {\text if}\,\, l=0 .
\end{array} 
\right.
\eea   
A  useful alternative representation of  modified Bessel's 
$K$ function which makes its properties  more transparent is given by
\be\label{modifiedB}
K_{i u}(y) = {1\over 2} \int^{\infty}_{-\infty} e^{- y \cosh t} e^{i u t}   dt. 
\ee
This expression allows analytical continuation of the wave function from the region $\Re\, s>1$ in (\ref{sum1}) into the whole complex plane $s$   because the   Bessel's  $K_s(y)$ functions are well defined for any $s$. 
Besides, an easy examination shows that the finite sum is:
\bea
\sum_{r=0}^{m-1}
e^{\frac{2\pi ilr}{m} }=\left\{
\begin{array}{l}
m\quad {\text if}\,\, m\,\, \text{divides}\,\, l\\
0\quad \text{otherwise}
\end{array}~~~ .
\right. 
\eea 
To summarise, for the Fourier coefficients (\ref{fcoefficients}) we shall get
\bea
c_l(y)=\frac{2 \pi^s}{\Gamma(s)}\,\tau_{s-\frac{1}{2}}(|l|)\sqrt{y} 
K_{s-\frac{1}{2}}(2 \pi  |l|y ), \quad {\text if}\,\, l\neq 0,
\eea
where
\bea
\tau_{\nu}(n)=\sum_{a \cdot b=n}\left(\frac{a}{b}\right)^\nu ,
\eea
while for $l=0$:
\bea
c_0(y)=\frac{\sqrt{\pi}\Gamma(s-\frac{1}{2})\zeta(2s-1)}
{\Gamma(s)}\,y^{1-s} .
\eea 
Thus we recovered the second solution $y^{1-s} $ in (\ref{solution3}) and calculated the coefficient 
$c^{'}_0$ in front of it.  Thus the invariant solution (\ref{solution4}) takes the following form:
\bea
\zeta(2s) \, \psi_{s}(x,y) &=& \zeta(2s) y^s + \frac{\sqrt{\pi}\Gamma(s-\frac{1}{2})\zeta(2s-1)}
{\Gamma(s)}\,y^{1-s} +\nn\\
&+&\sqrt{y} \frac{4 \pi^s}{\Gamma(s)} \sum_{l=1}^{\infty}\tau_{s-\frac{1}{2}}(l)
K_{s-\frac{1}{2}}(2 \pi  ly )\cos(2\pi l x) . \qquad 
\eea 
Using Riemann's reflection relation
\bea 
\zeta (s)=\frac{\pi ^{s-\frac{1}{2}} \Gamma \left(\frac{1-s}{2}\right)}
{ \Gamma \left(\frac{s}{2}\right)}\,\zeta (1-s)
\eea
and introducing the notation 
\bea 
\theta(s)=\pi ^{-s} \zeta (2 s) \Gamma (s)
\eea  
 we arrive at the elegant final expression for the energy eigenfunctions obtained by Maas \cite{maass}:
\bea\label{elegant}
\theta(s) \psi_{s}(z) &=&\theta (s)y^s+\theta(1-s)\,y^{1-s}
+4\sqrt{y} \sum_{l=1}^{\infty}\tau_{s-\frac{1}{2}}(l)
K_{s-\frac{1}{2}}(2 \pi  ly )\cos(2\pi l x) ,\qquad 
\eea 
This wave function is well defined in the complex $s$ plane and has a simple pole at $s=1$.
The physical continuous spectrum was defined in (\ref{srange}), where  $s=\frac{1}{2} +iu $, $u \in [0,\infty]$ so that 
\be\label{eigenvalues}
E= s(1-s)= \frac{1}{4} +u^2 .
\ee
The continuous spectrum wave functions $\psi_s(x,y)$ are delta function normalisable \cite{maass,roeleke,selberg1,selberg2,Faddeev,bump}.
The wave function (\ref{elegant}) can be  conveniently represented also in the form 
\bea\label{elegant1}
\psi_{ \frac{1}{2} +i u}(z) &=& y^{ \frac{1}{2} +i u}+{\theta(\frac{1}{2} -i u) \over \theta(\frac{1}{2} +i u)}   \,y^{\frac{1}{2} -i u}
+{4\sqrt{y} \over \theta(\frac{1}{2} +i u)}   \sum_{l=1}^{\infty}\tau_{i u}(l)
K_{i u }(2 \pi  ly )\cos(2\pi l x) , \nn\\
\eea
where 
\bea
K_{-i u}(y ) =K_{i u }( y),~~~~~\tau_{-i u}(l) =\tau_{i u}(l)~. 
\eea

The physical interpretation of the wave function becomes  more transparent when we introduce the new variables 
\be\label{newvariab}
\tilde{y} = \ln y,~~~~ p= -u,~~~~E=   p^2  + \frac{1}{4} 
\ee
as well as an alternative normalisation of the wave function:
\bea\label{alterwave}
&\psi_{p} (x,\tilde{y})  \equiv
y^{- \frac{1}{2}} \psi_{ \frac{1}{2} +i u}(z) =\\
&= e^{-i p \tilde{y}  }+{\theta(\frac{1}{2} +i p) \over \theta(\frac{1}{2} -i p)}   \, e^{  +i p \tilde{y}} + { 4  \over  \theta(\frac{1}{2} -i p)}  \sum_{l=1}^{\infty}\tau_{i p}(l)
K_{i p }(2 \pi  l e^{\tilde{y}} )\cos(2\pi l x) .\nn
\eea
Indeed, the first two terms describe the incoming and outgoing plane waves. The plane wave  $e^{-i p \tilde{y}  }$ incoming from infinity of the $y$ axis on Fig.\ref{fig1}-\ref{scattering}  ( the vertex $\CD$)  elastically scatters on the boundary $ACB$ of the fundamental region $\CF$ on Fig. \ref{fig1}.  The reflection amplitude is a pure phase and is given by the expression in front of the outgoing plane wave $e^{  i p \tilde{y}}$: 
\be\label{phase}
{\theta(\frac{1}{2} +i p) \over \theta(\frac{1}{2} -i p)} = \exp{[i\, \varphi(p)]}.
\ee
The rest of the wave function describes the standing waves $\cos(2\pi l x)$  in the $x$ direction between boundaries $x=\pm 1/2$
with the amplitudes $K_{i p }(2 \pi  l y )$, which are exponentially decreasing with index $l$.

In addition to the continuous spectrum the system (\ref{Laplace_eq})  has a discrete spectrum  
\cite{maass,roeleke,selberg1,selberg2,Faddeev,bump}.
The number of discrete states is infinite: $E_0=0 < E_1 < E_2 < ....\rightarrow \infty$, the spectrum is extended to infinity -  unbounded from above -  and lacks any accumulation  point  except infinity. 
Let us denote the wave functions of the discrete spectrum by $\psi_n(z)$ so that the expansion into the 
full set of basis vectors will take  the form 
\be
f(x,\tilde{y}) = \sum_{n \geq 0} a_n \, \psi_n(x,\tilde{y})   + {1\over 2 \pi }\int^{\infty}_{0} a_{p} \,  \psi_{p }(x,\tilde{y}) d p
\ee
and the Parseval identity will be
\bea
& \vert \vert f(z) \vert \vert^ 2= \sum_{n \geq 0} \vert a_n \vert^2  +  {1\over 2 \pi }\int^{\infty}_{0} \vert a_{p} \vert^2 d p ,
\eea
and 
\bea
& \sum_{n \geq 0}  \psi_n(x,\tilde{y}) \psi^*_n(x_1,\tilde{y}_1)  + {1\over 2 \pi }\int^{\infty}_{0} \psi_{p }(x,\tilde{y}) \psi_{-p }(x_1,\tilde{y}_1)  d p = \delta^{(2)}(z-z_1).\nn\\
\eea
The wave functions of the discrete spectrum have the following form \cite{maass,roeleke,selberg1,selberg2, winkler,hejhal,hejhal1}:
\bea\label{wavedisc}
\psi_n(z) &=&   \sum_{l=1}^{\infty} c_l(n) \,
\sqrt{y}\, K_{i u_n }(2 \pi  l y ) 
\{\begin{array}{ll} 
&\cos(2\pi l x) \\
&\sin(2\pi l x)   \\
\end{array}, 
\eea
where the spectrum $E_n = {1\over 4} + u^2_n$ and the coefficients $c_l(n)$  are not known analytically, but were computed numerically for many 
values of $n$ \cite{winkler,hejhal,hejhal1}.  Having explicit expressions of the wave functions one can analyse the quantum-mechanical  behaviour of the correlation functions, which we shall investigate in the next sections.

\section{\it Correlation functions }

\subsection{\it Two-point correlation function}

First let us calculate the two-point correlation function: 
\bea
\CD_2(\beta,t)=  \langle     A(t)   B(0) e^{-\beta H}   \rangle =
 \sum_{n}  \langle  n \vert e^{i H t } A(0) e^{-i H t } B(0) e^{-\beta H} \vert n  \rangle=\nn\\
 =\sum_{n,m} e^{i (E_n -E_m)t - \beta E_n}   \langle  n \vert   A(0)\vert m  \rangle  \langle m\vert  B(0)  \vert n  \rangle.
\eea
The energy eigenvalues (\ref{eigenvalues}) are parametrised by $n = {1\over 2} +i u$, $E_n =\frac{1}{4} + u^2 $ and $m = {1\over 2} +i v$, $E_m =\frac{1}{4} + v^2 $, thus 
\bea
\CD_2(\beta,t)=\int^{+\infty}_{0} \int^{+\infty}_{0} du \,  dv  ~ e^{i (u^2 -v^2)t - \beta( \frac{1}{4} + u^2)}  ~~~~~~~~~~~~ \nn \\
\int_{\CF}\psi_{ \frac{1}{2} -i u}(z)  \, A \, \psi_{ \frac{1}{2} +i v }(z) \,  d\mu(z)~
\int_{\CF}\psi_{ \frac{1}{2} -i v }(w)  \, B \, \psi_{ \frac{1}{2} +i u}(w) \,  d\mu(w)~,\nn\\
\eea
where the complex conjugate  function is $\psi^*_{ \frac{1}{2} +i u}(z) =\psi_{ \frac{1}{2} -i u}(z) $.
Defining the basic matrix element as
\bea\label{basicmatrix}
A_{uv} = 
\int_{\CF}  \psi_{ \frac{1}{2} -i u}(z)  \, A \, \psi_{ \frac{1}{2} +i v }(z) \,  d\mu(z)  =
\int^{1/2}_{-1/2} dx  \int^{\infty}_{\sqrt{1-x^2}} {dy \over y^2}  \psi_{ \frac{1}{2} -i u}(z)  \, A \, \psi_{ \frac{1}{2} +i v }(z)~~~~  
\eea
for the two-point correlation function we shall get 
\bea
\CD_2(\beta,t)=  \int^{+\infty}_{-\infty} e^{i (u^2 -v^2)t - \beta( \frac{1}{4} + u^2)}   
A_{uv}\,  B_{vu} \, du dv~.
\eea
In terms of the new variables (\ref{newvariab}) the  basic matrix element (\ref{basicmatrix}) will take the form 
\bea\label{basicmatrixelement}
A_{p q}  
= \int^{1/2}_{-1/2} dx  \int^{\infty}_{{1\over 2}\log(1-x^2)}  dy  ~ \psi^*_{p} (x,y)  \,  ( e^{- \frac{1}{2} y}   A \,   e^{  \frac{1}{2} y})  \, \psi_{ q} (x,y) \, . 
 \eea
The matrix element (\ref{basicmatrix}), (\ref{basicmatrixelement}) plays a fundamental role in the investigation of the correlation functions because all correlations can be expressed through it.  One should choose also appropriate observables 
$A$ and  $B$.  The operator $y^{-2} $  seems very appropriate for two reasons. Firstly, the convergence of the integrals over the fundamental region $\CF$ will be well defined. Secondly, this operator is reminiscent of 
the exponentiated Louiville  field  since $y^{-2} = e^{-2 \tilde{y}}$ . Thus we are interested in calculating  the matrix element  (\ref{basicmatrixelement})
for the observables in the form of the  Louiville-like operators:
\be\label{Louiville-like}
A(N)=  e^{-2 N y}. 
\ee  
We shall get\footnote{The other interesting observable is  $A = \cos(2\pi N x), N=1,2,... $.} 
\bea
A_{p q}(N) =\int^{1/2}_{-1/2} dx  \int^{\infty}_{{1\over 2}\log(1-x^2)}  dy  ~ \psi^*_{p} (x,y)  \,   e^{- 2 N y}  \, \psi_{ q} (x,y) \,  ~,~~ N=1,2,...
 \eea
Calculating the above matrix elements we shall use a perturbative expansion in which the part of the wave function (\ref{alterwave}) containing the Bessel's functions and the contribution of the discrete spectrum (\ref{wavedisc}) will be considered as a perturbation. As we shall demonstrate below, these terms of the perturbative expansion are small and don't influence our results.  The reason behind this fact is that  in the integration region $\Im z \gg 1, \Im w \gg 1$ of the matrix element (\ref{basicmatrix}) the Bessel's functions  decay exponentially, as one can get convinced by inspecting (\ref{modifiedB}). Therefore the contribution of these high modes is small (analogues to the so called mini-superspace approximation in the Liouville theory).  Thus we shall consider the perturbative expansion over high frequency modes $l=1,2,...$ in (\ref{alterwave}) and (\ref{wavedisc}). In the first approximation of the wave function (\ref{alterwave}) for the matrix element we shall get 
\bea
 A_{p q}(N)   
 &=& \int^{1/2}_{-1/2} d x 
 \left(   {  (1-x^2)^{-N+ {p-q \over 2 i}}       \over  2 N+i (p-q) }     +
  {  (1-x^2)^{-N+ {p+q \over 2 i}}  e^{-i \varphi(q)}     \over  2 N+i(p+q)  }  \right.\nn\\
  \nn\\
&& \left.   +
  {  (1-x^2)^{-N- {p+q \over 2 i}}  e^{i \varphi(p)}     \over  2 N -i (p+q)  }   
  +  {  (1-x^2)^{-N- {p-q \over 2 i}}    e^{i ( \varphi(p)-\varphi(q)) }    \over  2 N-i (p-q)  }                       \right) , 
\eea
where we used  (\ref{phase}).  
Integration over $x$ can be performed exactly with the result for the basic matrix element of the following form:
\bea
 A_{pq}(N)= {\, _2F_1 \left(  \frac{1}{2}, N+  i {p-q \over 2 }  ;    \frac{3}{2};   \frac{1}{4}          \right)  \over   2 N +i (p-q)} +{\, _2F_1 \left(  \frac{1}{2}, N+i {p+q \over 2 }  ;    \frac{3}{2};   \frac{1}{4}          \right)  \over   2 N+ i (p+q)} e^{-i \varphi(q)} + \nn\\
 +{\, _2F_1 \left(  \frac{1}{2}, N- i {p+q \over 2 }  ;    \frac{3}{2};   \frac{1}{4}          \right)  \over   2 N -i (p+q) } e^{i \varphi(p)} +   {\, _2F_1 \left(  \frac{1}{2}, N-i {p-q \over 2 }  ;    \frac{3}{2};   \frac{1}{4}          \right)  \over   2 N+ i (p-q )} e^{i (\varphi(p)-\varphi(q))}, 
\eea
where the reflation phase $\varphi(p)$ was defined in (\ref{phase}). Thus for the two-point correlation function we shall get
\bea\label{twopointcorrel}
\CD_2(\beta,t)=  \int^{+\infty}_{-\infty} e^{i (p^2 -q^2)t - \beta( \frac{1}{4} + p^2)}   
A_{pq}(N)\,  A_{qp}(M) \, dp dq~.
\eea
 \begin{figure}[h]
 \centering
        \includegraphics[width=0.3\textwidth]{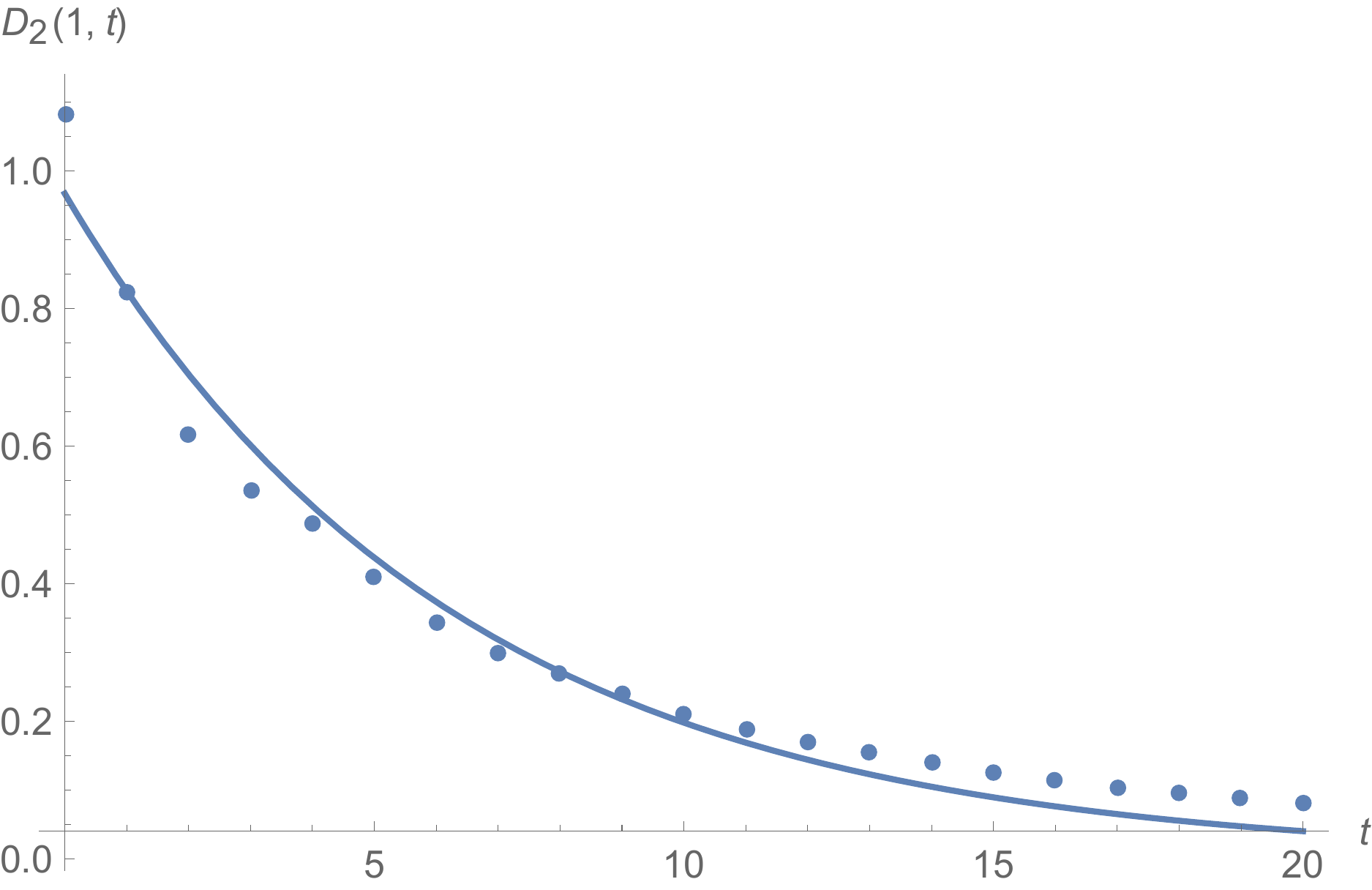}~~~~
         \includegraphics[width=0.3\textwidth]{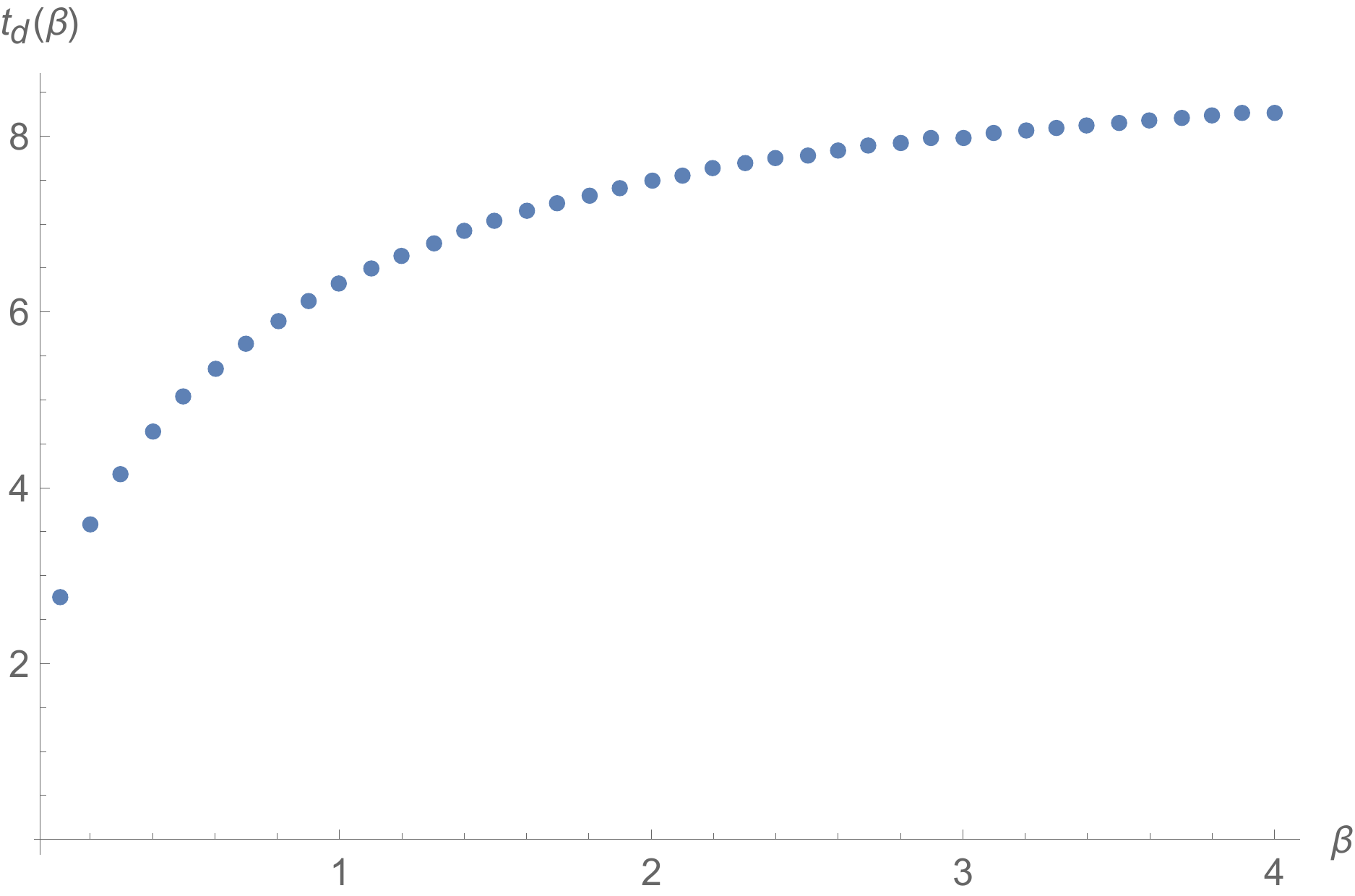}~~~~
         \includegraphics[width=0.3\textwidth]{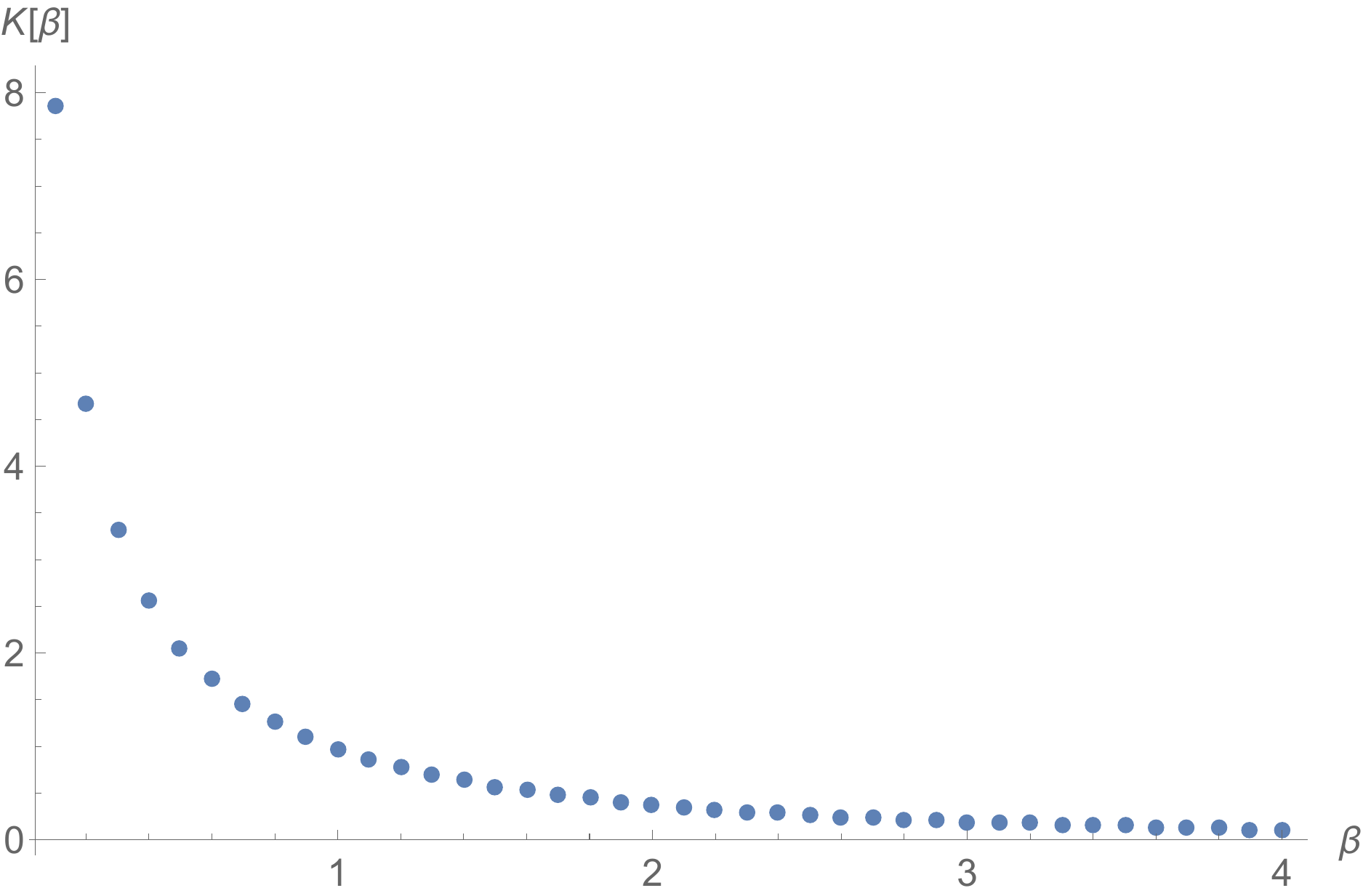}
        \caption{The exponential decay of the two-point correlation function $\CD_2(\beta,t)$ as a function of time  at   temperature $\beta =1$.  The points are fitted by the curve $K(\beta) \exp{(- t / t_d(\beta))}$. The exponent $t_d(\beta)$ has a well defined high and low temperature limits. The limiting values in dimensionless units are $t_d(0) \approx 0.276$ and $t_d(\infty) \approx 0.749$. The temperature dependence of $K(\beta)$ is shown on the l.h.s. graph.
}
\label{twopointfunc}
\end{figure}
The correlation function is between two Louiville-like  fields in the power $N$ and $M$ respectively.
This expression is very convenient for the analytical and numerical analyses.   
It is expected that the two-point correlation function should decay exponentially  \cite{Maldacena:2015waa},
therefore we fitted the points by the curve  
\be
\CD_2(\beta,t)  \sim K(\beta)~e^{-{t \over t_d(\beta)}}.
\ee
The $ t_d(\beta)$ is the decorrelation time and it defines one of the characteristic time scales in the quantum-mechanical system. The exponential decay of the two-point correlation function with time at different  temperatures is shown on Fig.\ref{twopointfunc}\footnote{We would like to thank Gabriel Poghosyan for numerical calculation of the two-point correlation function presented on Fig.\ref{twopointfunc}.} .
The dependence of the exponent $ t_d(\beta)$ and of the prefactor $K(\beta)$ as a function of temperature are presented in Fig.\ref{twopointfunc}.   
 As one can see, at high and low temperatures the decorrelation time tends to the fixed values. These limiting values we calculated by using the expressions (\ref{twopointcorrel1}) and (\ref{twopointcorrel}). At large $\beta \rightarrow \infty$ the main contribution came from the zero momentum region $p=0$:  
 \bea\label{twopointcorrel1}
\CD_2(\infty,t)=  \int^{+\infty}_{-\infty} e^{ -i q^2 t  }   
A_{0q}(N)\,  A_{q0}(M) \, dq~\sim K(\infty)~e^{-{t \over t_d(\infty)}},
\eea
and at $\beta \rightarrow 0$  from the region $p=q$: 
\bea\label{twopointcorrel}
\CD_2(0,t)=  \int^{+\infty}_{-\infty} e^{i (p^2 -q^2)t  }   
A_{pq}(N)\,  A_{qp}(M) \, dp dq~\sim K(0)~e^{-{t \over t_d(0)}}.
\eea
The corresponding limiting values in dimensionless units are  shown on the l.h.s. of the Fig.\ref{twopointfunc}.

\subsection{\it Four-point correlation function}

It was conjectured in the literature \cite{Maldacena:2015waa} that the classical chaos can be diagnosed in thermal quantum systems by using an out-of-time-order correlation functions as well as by the square of the commutator of the operators which are separated in time. The out-of-time four-point correlation function of interest was defined in \cite{Maldacena:2015waa} as follows:
\bea\label{outoftime}
\CD_4(\beta,t)=  \langle   A(t)   B(0) A(t)   B(0)e^{-\beta H}   \rangle= ~~~~~~~~~~~~~~~~~~~~~~~~~~~~~~\\
 =\sum_{n,m,l,r} e^{i (E_n -E_m+E_l - E_r)t - \beta E_n}  \langle  n \vert   A(0)\vert m \rangle \langle m\vert  B(0)   \vert l \rangle
\langle l \vert   A(0)\vert r \rangle  \langle r\vert  B(0)   \vert n \rangle.   \nn
\eea
The other important observable which we shall consider here is the square of the commutator of the Louiville-like  operators  separated in time \cite{Maldacena:2015waa} 
\be\label{commutatorL}
C(\beta,t) = \langle [A(t),B(0)]^2 e^{-\beta H} \rangle~.
\ee 
The energy eigenvalues we shall parametrise as $n = {1\over 2} +i u$, $m = {1\over 2} +i v$,$l = {1\over 2} +i l$ and $r = {1\over 2} +i r$, thus  from (\ref{outoftime}) we shall get
\bea\label{fourmatrix}
\CD_4(\beta,t)&=& \int^{+\infty}_{-\infty} e^{i (u^2 -v^2 + l^2  - r^2)t - \beta( \frac{1}{4} + u^2)}  ~ 
A_{uv}\,  B_{vl} \, A_{lr}\,  B_{ru} \,  du dv dl dr ~. 
 \eea
 \begin{figure}
 \centering
 \includegraphics[angle=0,width=6cm]{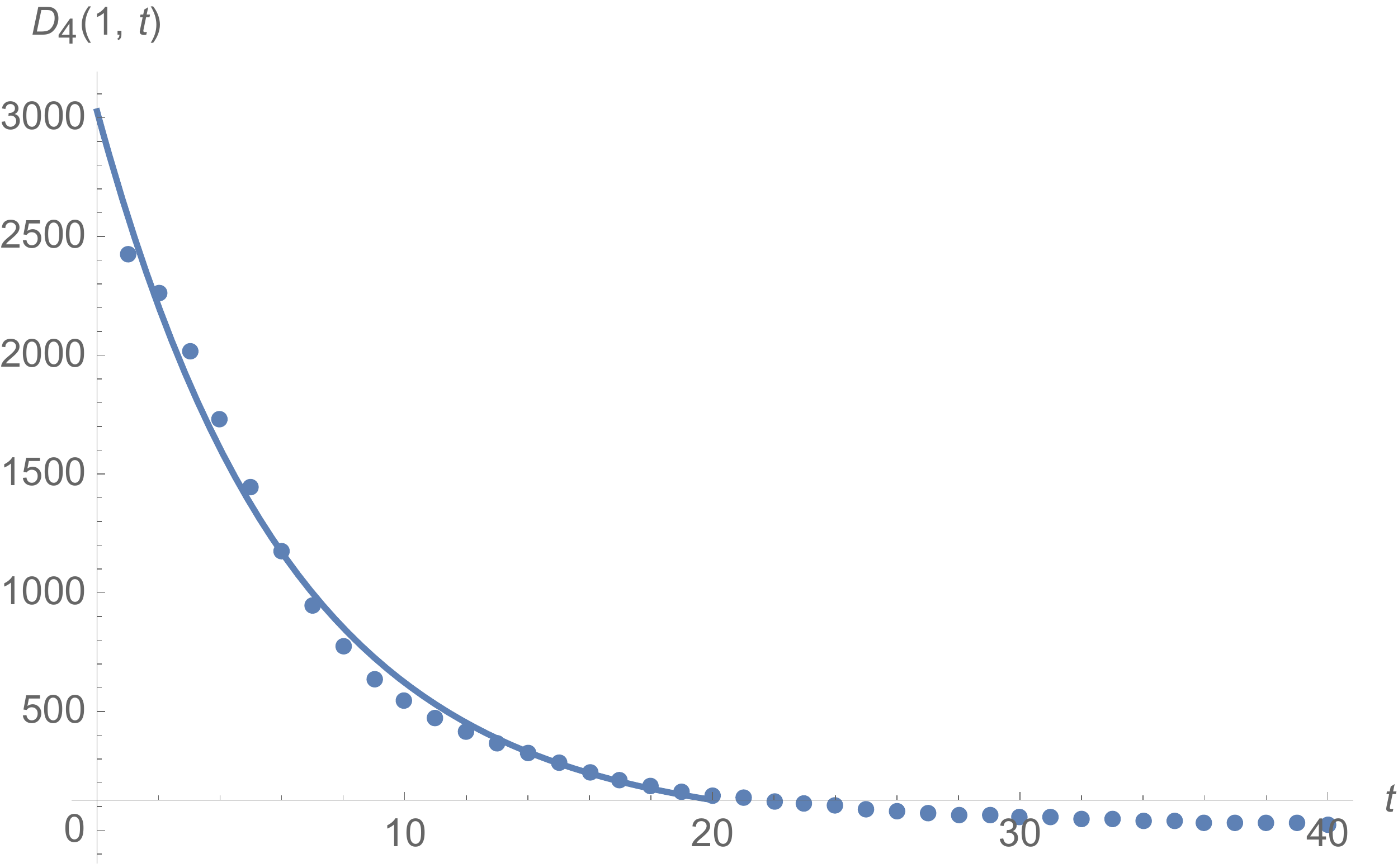}~~~~~~~~~
  \includegraphics[angle=0,width=5cm]{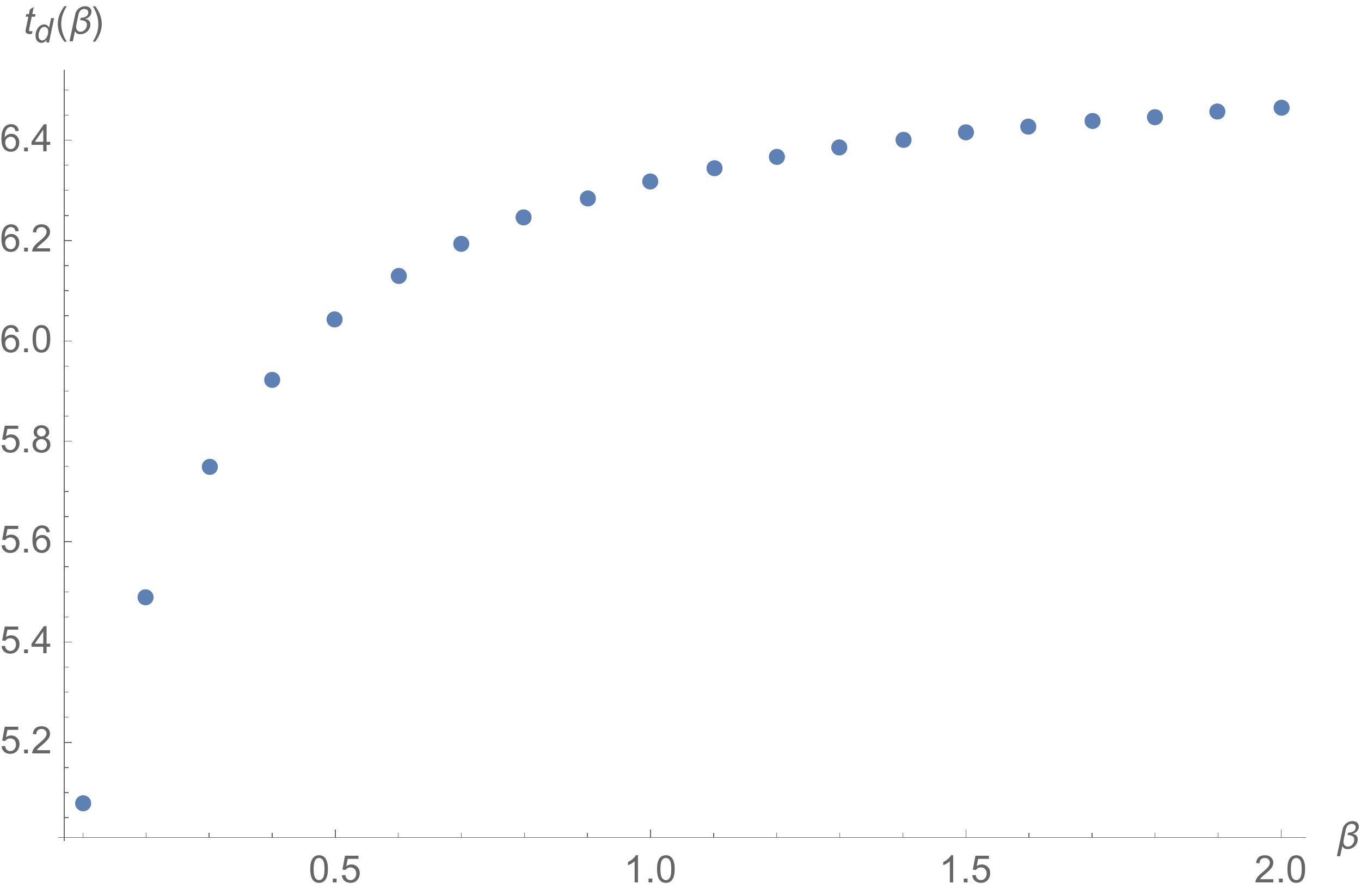}
\caption{  The exponential decay of the correlation function $\CD_4(\beta,t)$ as a function of time at $\beta =1$. The rest of the functions $\CD'_4(\beta,t),\CD''_4(\beta,t),\CD'''_4(\beta,t)$ demonstrate a similar exponential  decay $  \sim ~\exp{(-{t \over t_d(\beta)} )}$. The temperature dependence of the exponent $t_d(\beta)$ have a well defined high and low temperature limits and is shown on the r.h.s. graph. 
The corresponding limiting values of the function $t_d(\beta)$ in dimensionless units are $t_d(0)=0,112$ and $t_d(\infty)=0,163$. The behaviour of the exponent $t_d(\beta)$ of the two-point correlation function is shown on the Fig.\ref{twopointfunc}.
}
\label{fourpointfunc}
\end{figure}
In terms of the variables (\ref{newvariab})   the four-point correlation function (\ref{fourmatrix})  will take the following form: 
\bea
\CD_4(\beta,t)=  \int^{+\infty}_{-\infty} e^{i (p^2 -q^2 + l^2  - r^2)t - \beta( \frac{1}{4} + p^2)}  ~ 
A_{pq}(N)\,  A_{ql}(M)\,  A_{lr}(N)\,  A_{rp}(M) \,   dp dq  dl dr .  \nn\\
\eea
As it was suggested in \cite{Maldacena:2015waa}, the most important correlation function indicating the traces of the classical chaotic dynamics in quantum regime is (\ref{commutatorL}) 
\bea\label{commutatorL1}
C(\beta,t) = - \CD_4(\beta,t) + \CD'_4(\beta,t) +\CD''_4(\beta,t)-\CD'''_4(\beta,t), 
\eea
where in the case of the Artin system  we shall get  (see Appendix for details)
\bea\label{fourpointfunct1}
&\CD'_4(\beta,t) +  \CD''_4(\beta,t) =  \\
& 2 \int^{+\infty}_{-\infty} e^{- \beta( \frac{1}{4} + p^2)}  ~\cos{ ( q^2 - r^2)t } ~ 
A_{pq}(N)\,  A_{ql}(M)\,  A_{lr}(N)\,  A_{rp}(M) \,   dp dq  dl dr,   \nn
\eea
and 
\bea\label{fourpointfunct2}
&\CD_4(\beta,t) +  \CD'''_4(\beta,t) =  \\
& 2 \int^{+\infty}_{-\infty} e^{- \beta( \frac{1}{4} + p^2)}  ~\cos{(p^2 -q^2 + l^2  - r^2)t } ~ 
A_{pq}(N)\,  A_{ql}(M)\,  A_{lr}(N)\,  A_{rp}(M) \,   dp dq  dl dr  . \nn
\eea
On Fig.\ref{fourpointfunc} one can see the behaviour of the four-point correlations as the function of the temperature and time. All four correlation functions decay exponentially: 
\be
\CD_4(\beta,t)  \sim K(\beta)~e^{-{t \over t_d(\beta)}}.
\ee
\begin{figure}
 \centering
 \includegraphics[angle=0,width=5cm]{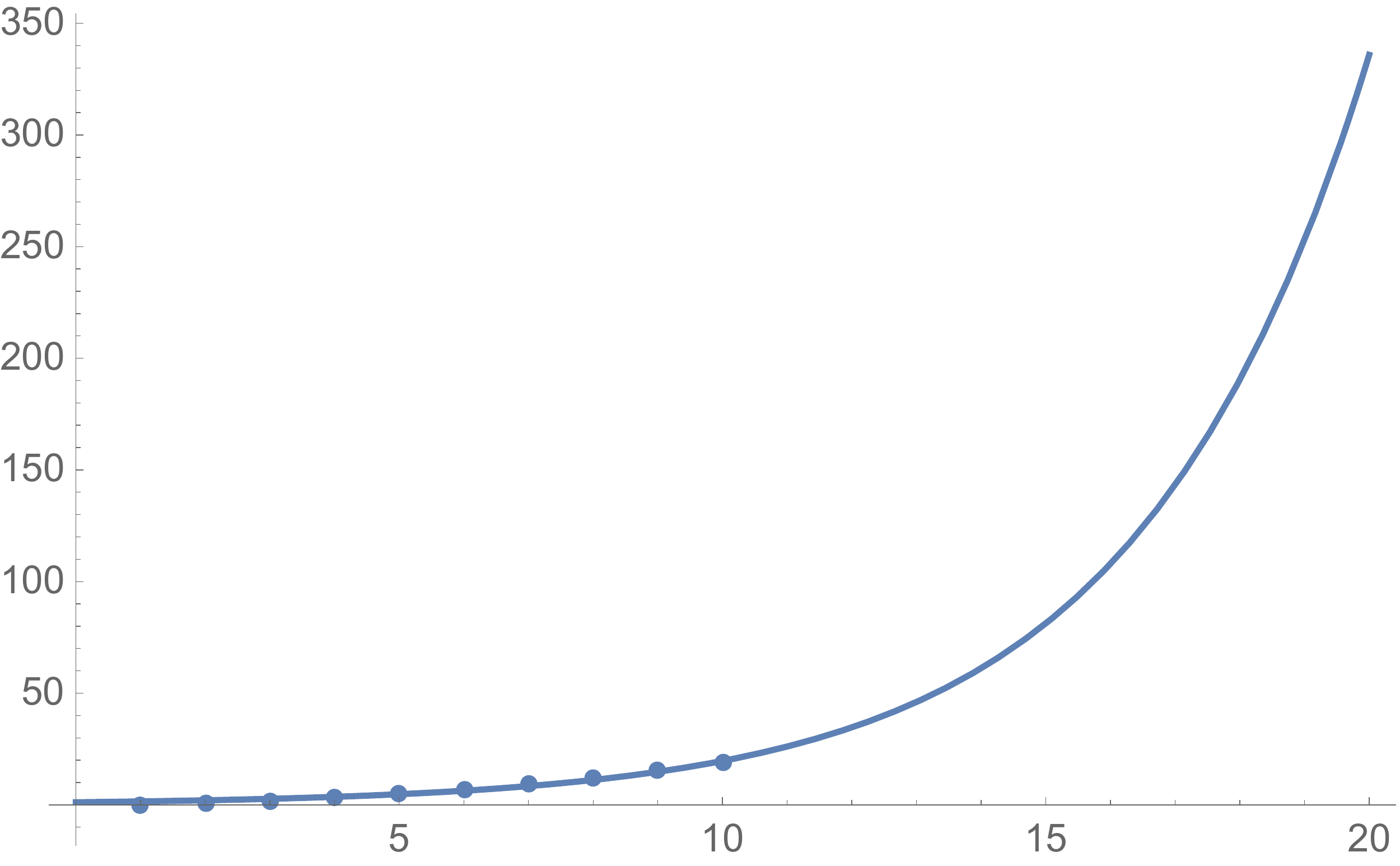}~~~~~~
  \includegraphics[angle=0,width=7cm]{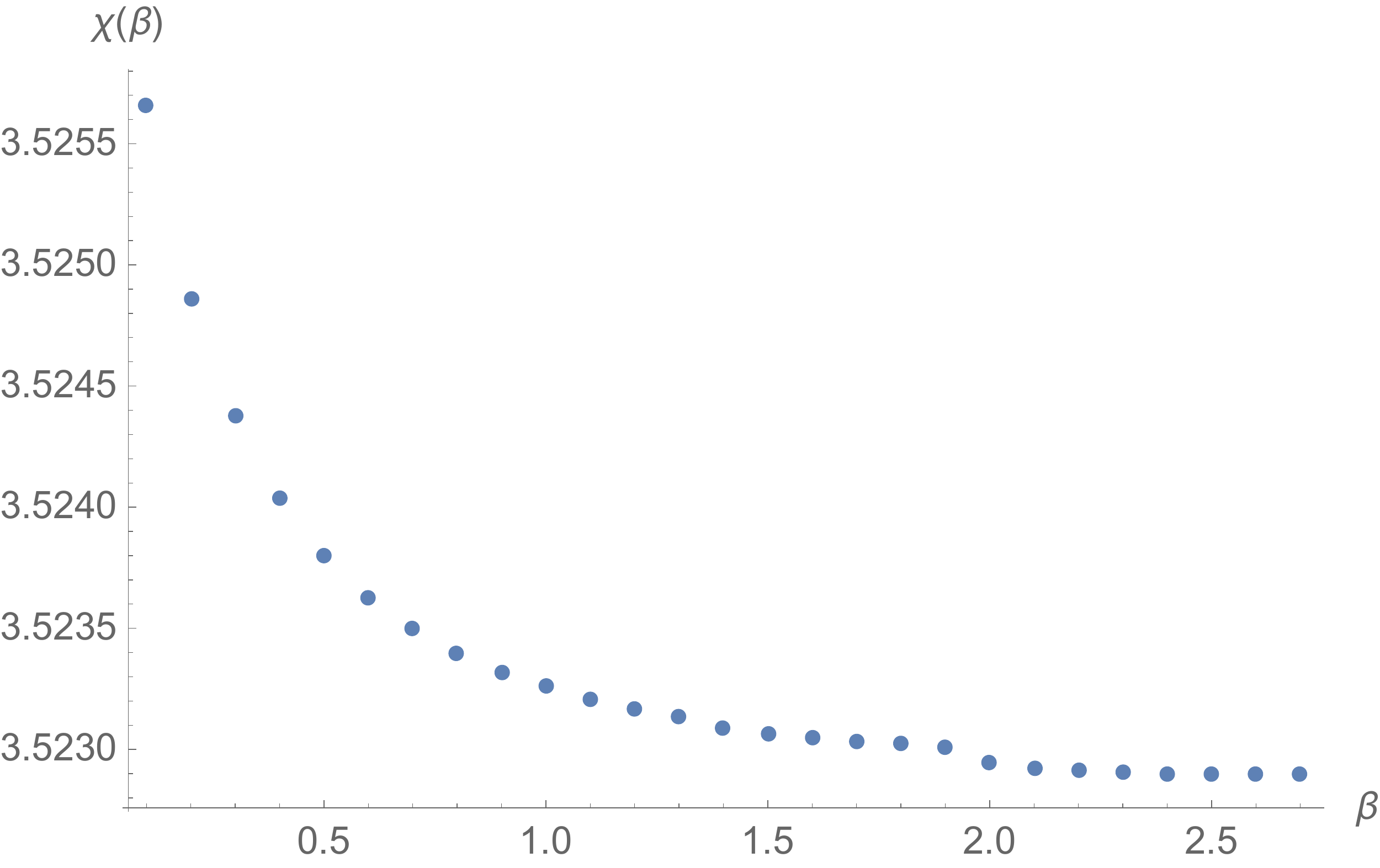}
\caption{  Time evolution of the correlation function $C(\beta,t)$ (\ref{commuta}) at temperature $\beta =0.1$.  For the short  time intervals the function $C(\beta,t)$ exponentially increases with time. This behaviour is reminiscent to the exponential divergency of the classical  trajectories in hyperbolic dynamical systems.  As one can see, the exponent  $\chi(\beta)$ which defines the behaviour of the correlation function of the operators separated in time in  the commutator (\ref{commuta}), (\ref{exponentc1}) slowly decreases with $\beta$. Such behaviour of the correlation function $C(\beta,t)$ does not saturate the maximal growth future (\ref{lineargrowth}) which should be linear in $\beta$ (\ref{linear1}). 
}
\label{timeevolutionofcommutator}
\end{figure}
Now we can turn to the investigation of the commutator (\ref{commutatorL}), (\ref{commutatorL1}) which can be represented in the following form:  
\bea\label{commuta}
C(\beta,t) = 2 \int^{+\infty}_{-\infty} e^{- \beta( \frac{1}{4} + p^2)}\{ \cos{ ( q^2 - r^2)t }  - \cos{ (p^2 -q^2 + l^2  - r^2)t }  \}\nn\\
A_{pq}(N)\,  A_{ql}(M)\,  A_{lr}(N)\,  A_{rp}(M) \,   dp dq  dl dr  ,\nn\\
=  \int^{+\infty}_{-\infty} e^{- \beta( \frac{1}{4} + p^2)}   \sin{ {1\over 2}( p^2 + l^2 - 2 r^2)t } ~\cdot~ \sin{ {1\over 2}( p^2 + l^2 - 2 q^2)t }  \nn\\
A_{pq}(N)\,  A_{ql}(M)\,  A_{lr}(N)\,  A_{rp}(M) \,   dp dq  dl dr  ,
\eea
where we used  (\ref{fourpointfunct1}) and (\ref{fourpointfunct2}).  It was conjectured in \cite{Maldacena:2015waa}  that the influence of chaos on the commutator $C(\beta,t)$  can develop no faster than exponentially: 
\be \label{lineargrowth}
C(\beta,t) \approx f(\beta)\, e^{{2\pi  \over \beta} t} ,~~%~~\lambda={2\pi  \over \beta} = f(T)\, e^{ 2\pi T t } 
\ee
with the exponent ${2\pi  \over \beta}t = 2\pi T t  $ which is linear in temperature $1/ \beta =T$ and time $t$. 
Calculating the function $C(\beta,t)$ one can check if it  grows exponentially: 
\be\label{exponentc1}
C(\beta,t) \sim f(\beta) \,e^{ {2\pi  \over  \chi(\beta)} t }, 
\ee
and if the exponent $\chi(\beta)$ is linear in $\beta$.  The maximal growth will correspond  to the exponent $\chi(\beta)$ which grows linearly with $\beta$  (\ref{lineargrowth}), that is, 
\be\label{linear1}
\chi(\beta) \sim \beta.
\ee
The results of the integration are presented on the Fig.\ref{timeevolutionofcommutator}. This beautifully confirms the fact that the correlation function $C(\beta,t)$ indeed grows exponentially with time  as it takes place in its classical counterpart. As one can see, the exponent  $\chi(\beta)$ defining the behaviour of the commutator $C(\beta,t)$   in (\ref{commuta}) and (\ref{exponentc1}) slowly decreases with $\beta$. Such behaviour of the commutator $C(\beta,t)$ does not saturate the maximal growth of the correlation function (\ref{linear1}) which should be linear in $\beta$.

In order to check if the results are sensitive to the truncation of the high modes of the Maass wave function (\ref{elegant}) we included their contribution into the integration of the  basic matrix element $A_{uv}$  in (\ref{basicmatrix}).  We found that  their influence on the behaviour of the correlation functions is negligible. The numerical values of the exponents are changing in the range of few percentage and do not influence the results.  In summary,  all two and four-point correlation functions decay exponentially.   The commutator $C(\beta,t)$ in (\ref{commuta}) and (\ref{exponentc1}) grows exponentially with exponent which is almost constant Fig.\ref{timeevolutionofcommutator}. This behaviour does not saturate the condition of the maximal growth (\ref{lineargrowth}), (\ref{linear1} ).

\section{\it Acknowledgment }

R.P. is thankful to the NCSR"Demokritos" for hospitality during June-July 2018 when this work was completed. H.B. would like to thank NCSR"Demokritos" for kind hospitality and generous financial support of the MIXMAX project.  G.S. would like to thank  Luis Alvarez-Gaume for stimulating discussions of $SL(2,Z)$ automorphic functions, the references \cite{bump,hejhal2} and kind hospitality at the Simon Center for Geometry and Physics where part of this work was completed.  This project received funding from the European Union's Horizon 2020 research and innovation programme under the Marie Sk\'lodowska-Curie grant agreement No 644121. 

\section{\it Appendix } 
In order to compute the square of the commutator  (\ref{commuta})
one should consider the following four-point correlation functions 
\bea
\CD'_4(\beta,t)= <   A(t)   B(0)  B(0) A(t)   e^{-\beta H}   >= ~~~~~~~~~~~~~~~~~~~~~~~~~~~~~~\\
 =\sum_{n,m,l,r} e^{i ( -E_m + E_r)t - \beta E_n}  < n \vert   A(0)\vert m> <m\vert  B(0)   \vert l>
< l \vert   B(0)\vert r> <r\vert  A(0)   \vert n>   \nn
\eea
\bea
\CD''_4(\beta,t)=  <   B(0) A(t)    A(t)   B(0)e^{-\beta H}   >= ~~~~~~~~~~~~~~~~~~~~~~~~~~~~~~\\
 =\sum_{n,m,l,r} e^{i ( E_m - E_r)t - \beta E_n}  < n \vert   B(0)\vert m> <m\vert  A(0)   \vert l>
< l \vert   A(0)\vert r> <r\vert  B(0)   \vert n>   \nn
\eea
\bea
\CD'''_4(\beta,t)=  <    B(0)  A(t)   B(0)  A(t)  e^{-\beta H}   >= ~~~~~~~~~~~~~~~~~~~~~~~~~~~~~~\\
 =\sum_{n,m,l,r} e^{i (-E_n +E_m-E_l + E_r)t - \beta E_n}  < n \vert   B(0)\vert m> <m\vert  A(0)   \vert l>
< l \vert   B(0)\vert r> <r\vert  A(0)   \vert n> .  \nn
\eea
These  correlation functions in the momentum space representation are given in (\ref{fourpointfunct1}) and (\ref{fourpointfunct2}).

\end{document}